\documentstyle[12pt]{article}
\def\l{\lambda}
\input{epsf}
\catcode`\@=11
\newcount\hour
\newcount\minute
\newtoks\amorpm
\hour=\time\divide\hour by60
\minute=\time{\multiply\hour by60 \global\advance\minute by-\hour}
\edef\standardtime{{\ifnum\hour<12 \global\amorpm={am}%
        \else\global\amorpm={pm}\advance\hour by-12 \fi
        \ifnum\hour=0 \hour=12 \fi
        \number\hour:\ifnum\minute<10 0\fi\number\minute\the\amorpm}}
\edef\militarytime{\number\hour:\ifnum\minute<10 0\fi\number\minute}
\def\draftlabel#1{{\@bsphack\if@filesw {\let\thepage\relax
   \xdef\@gtempa{\write\@auxout{\string
      \newlabel{#1}{{\@currentlabel}{\thepage}}}}}\@gtempa
   \if@nobreak \ifvmode\nobreak\fi\fi\fi\@esphack}
        \gdef\@eqnlabel{#1}}
\def\@eqnlabel{}
\def\@vacuum{}
\def\draftmarginnote#1{\marginpar{\raggedright\scriptsize\tt#1}}
\def\draft{\oddsidemargin -.2truein
        \def\@oddfoot{\sl preliminary draft \hfil
        \rm\thepage\hfil\sl\today\quad\militarytime}
        \let\@evenfoot\@oddfoot \overfullrule 3pt
        \let\label=\draftlabel
        \let\marginnote=\draftmarginnote
   \def\@eqnnum{(\theequation)\rlap{\kern\marginparsep\tt\@eqnlabel}%
\global\let\@eqnlabel\@vacuum}  }
\relax
\def\ads{AdS}

\renewcommand{\theequation}{\arabic{section}.\arabic{equation}}

\def\be{\begin{equation}}
\def\ee{\end{equation}}
\def\bs{\begin{subequations}}
\def\es{\end{subequations}}

\newcommand{\prl}{Phys.\ Rev.\ Lett.~}
\newcommand{\pr}{Phys.\ Rev.\ D~}

\newcommand{\np}{Nucl.\ Phys.\ B~}
\def\g{\gamma}

\begin{document}

\begin{titlepage}
\begin{flushright}
hep-th/9906206
\end{flushright}
\vspace{.0in}

\begin{center}
{\Large\bf   Supergravity, D-brane Probes and thermal super Yang-Mills:
a comparison\footnote{\small Research
supported in part by the EEC
under the TMR contract
ERBFMRX-CT96-0090.}}
\vskip .5cm

{\large Elias Kiritsis}
\vspace{.5in}

{\em
Physics Department, University of Crete\\
71003, Heraklion, GREECE.}

\end{center}

\vskip .45in

\begin{center} {\bf ABSTRACT }
 \end{center}
\begin{quotation}\noindent

A D3-brane probe in the context of AdS/CFT correspondence at finite
temperature is considered.
The supergravity predictions for the physical
effective  couplings of the
world-volume gauge theory of the probe brane are compared
to those calculated in one-loop perturbation theory in the
thermal gauge theory.
It is  argued that when the Higgs expectation value is much larger than
the
temperature, the supergravity result must agree with perturbative
thermal
Yang-Mills.
This provides a perturbative test of the Maldacena conjecture.
Predictions for the running electric and magnetic
effective couplings, beyond perturbation theory are also obtained.
Phenomenological applications for universe-branes
are discussed.
In particular mechanisms are suggested for reducing the induced
cosmological constant and naturally obtaining a varying speed of light
and a consequent inflation on the universe brane.

\end{quotation}
\vskip 1.5cm

\begin{flushleft}
June 1999
\end{flushleft}
\end{titlepage}
\vfill
\eject

\def\baselinestretch{1.2}
\baselineskip 14 pt

\noindent

\setcounter{section}{0}
\section{Introduction}

The central elements of
CFT/anti-De Sitter (aDS) correspondence \cite{mald,gkp,w1} are
the black D-brane solutions of type II supergravity \cite{hs} and
their
near-horizon geometry \cite{gt} along with their microscopic
interpretation
\cite{pol}.
In particular, the (3+1)-dimensional world-volume of
$N$ coinciding, extremal D3-branes is the arena of ${\cal N}{=}4$
supersymmetric $SU(N)$ Yang-Mills (SYM) theory which in the large $N$
limit,
according to the Maldacena conjecture \cite{mald},
is dual to type IIB superstrings propagating on the
near-horizon $\ads_5\times S^5$ background geometry.
There is a further proposal  \cite{w2}  linking the
thermodynamics  of large $N$, ${\cal N}{=}4$ supersymmetric
$SU(N)$ Yang-Mills  theory with the thermodynamics of Schwarzschild
black holes embedded in the AdS space \cite{hp}.
The classical geometry of black holes with Hawking temperature $T$
encodes the magnetic confinement, mass gap and other qualitative
features of large $N$ gauge theory heated
up to the same temperature. At the computational level,
the quantity that has been discussed to the largest
extent \cite{gkpeet,w2,gkt}-\cite{ft} is the Bekenstein-Hawking
entropy
which, in the Maldacena limit,
should be related to the entropy of Yang-Mills gas at
$N\rightarrow\infty$ and large 't Hooft coupling $g^2_{\rm YM}N$.

Turning-on Higgs expectation values in the gauge theory corresponds in
the
brane picture to moving around the D3-branes.
A useful configuration is one in which one (or a few) D-branes
are put a distance $r$ away from the stack of N D-branes (while being
kept
parallel).
In gauge theory this amounts to turning on a Higgs expectation value
$r/\alpha'$ breaking the gauge symmetry $SU(N+1)\to SU(N)\times U(1)$.

In the limit of $N\to \infty$  and large 't Hooft coupling $\l$,
the D-brane stack is well
described by supergravity with a specific background field
configuration.
Thus, the world-volume action of the probe brane can be evaluated
from the knowledge of the coupling of the world-volume gauge theory to
the bulk
supergravity fields (see for example \cite{db}).
On the other hand at small $\l$ the effective action of the probe brane
can be
computed in Yang-Mills perturbation theory.
At extremality (zero temperature in the gauge theory) ${\cal N}=4$
supersymmetry is known to prohibit renormalization of the
two-derivative
effective action and this is also visible in the supergravity
prediction.
There are renormalizations to the four-derivative effective action
and the leading supergravity term is expected to be given by the
one-loop
gauge theory  result \cite{ds,li}.

Supersymmetry can be broken softly by putting the gauge theory in a
heat bath
with temperature $T$.
This is expected to be described in supergravity by the near horizon
limit of a
black D-brane, namely an AdS black-hole  whose
Hawking temperature is $T$ \cite{w2}.
Moreover, a probe brane sitting outside of the horizon at a distance r
is described by the U(1) part of thermal ${\cal N}=4$ SYM $SU(N+1)\to
SU(N)\times U(1)$.
Several phenomena as well as the fate of the original scale duality can
be still
studied in this context \cite{doug}-\cite{1}.

Here we will study the effective gauge theory action on the probe
D3-brane
at finite temperature.
At strong 't Hooft coupling $\l$, it will be obtained from
supergravity.
The relevant data are the classical black-brane solution as well as the
world-volume D-brane action.
The relevant scales appearing in the theory are the Higgs expectation
value
$u$  and the temperature $T$ (or rather the thermal wavelength
$\sqrt{\l}~T$.
{}From the supergravity point of view, $\pi\sqrt{\l}~T\leq u$. At
$\pi\sqrt{\l}~T= u$ corresponds the black-brane horizon.
Thus, from supergravity we can calculate the effective potential, the
kinetic
terms as well as higher derivative terms (for example $F^4$ terms).
This effective action is supposed to be valid as $N\to \infty$ and
$\lambda$
large.

In perturbation theory the  relevant diagrams are
open string diagrams with a number of boundaries
attached to the stack of N D3-branes and a single boundary stack on the
probe
brane \cite{mal}.
Their $\alpha'\to 0$ limit provides the appropriate gauge theory
diagrams.
They integrate out the SU(N) and massive degrees of freedom.

We argue here, that in the supersymmetry restoration limit
$\sqrt{\l}~T<< u$
the supergravity result has a natural expansion in powers of $\l$, and
can
be compared with perturbative Yang-Mills calculations.
A similar situation occurs in near extremal calculations of black-hole
gray-body factors \cite{gray}.
The key ingredient for such behavior is supersymmetric
non-renormalization
theorems valid in the limit where supersymmetry is restored
\cite{nonr}.

Supergravity predicts that the leading contribution to the effective
potential
comes from three-loops and is proportional to $T^8/u^4$.
As a first test we do the one-loop computation of the vacuum energy by
integrating-out the massive open strings, and we finally take the
$\alpha'\to 0$ limit. In the relevant supersymmetric limit
$\sqrt{\l}~T<< u$
the one-loop contribution is exponentially suppressed as $e^{-u/T}$.
This contribution can be thought-of a non-perturbative contribution
due to a
solitonic string (see also \cite{pol2}).
For this to be true we must have the following hierarchy:
$\sqrt{\l}<<T/u<<1$.
Thus, up to exponentially suppressed contributions the one-loop
calculation
gives zero for the potential in the limit $T/u<<1$ in accordance with
supergravity.
The exponentially suppressed contributions will be absent if we treat
the fundamental multiplet with mass $\sim u$ (corresponding to the open
string stretched between black and probe brane) to be at zero temperature.
We can thus formulate the following:

$\bullet$ {\bf Conjecture:} Consider ${\cal N}=4$ superYM $SU(N+M)\to
SU(N)\times SU(M)\times U(1)$ by a Higgs expectation value $u$, 't
Hooft coupling $\l$
 and $N>>M$.
Consider also that the SU(N) part is in a heat bath with temperature T.
For $\l~T<< u$ the ${\cal O}(N)$ part of the vacuum energy has a
leading behavior $\sim \l^2 T^8/u^4$ and is given by supergravity.
The  ${\cal O}(N)$ part of the vacuum energy can be considered as the
vacuum
energy induced on the probe brane due to quantum effects on the other
brane or the bulk.

A similar investigation of the kinetic terms for scalars and gauge
bosons
gives a supergravity derived result which starts at two loops in the
limit
$T/u<<1$.
A  one-loop calculation gives an exponentially suppressed result (or
zero for non-thermal fundamental)
, in
accordance with the supergravity calculation.
Finally, supergravity predicts a one-loop (and higher) contribution to
the
$F^4$ couplings of the probe brane.
The one-loop open-string/gauge theory calculation is performed and
gives
agreement with supergravity as expected.

There are two phenomena observed here that may be potentially important
for
phenomenological purposes.

$\bullet$ \underline{Suppression of vacuum energy on a brane}.
It is popular lately to consider our four dimensional universe as a
three-brane,
embedded in ten dimensional, (partially) compactified spacetime.
Moreover, other three-branes may be providing mirror universes.
The prototype of this  is the Ho\v rava-Witten interpretation of the
Heterotic String \cite{hw}.
One of the important effects in this context is supersymmetry breaking
in a mirror brane and its communication in our universe.
Here we have a toy model of this situation.
In the presence of supersymmetry, the spectator brane is the
black-brane
while our universe is represented by the probe brane.
Both three-branes carry ${\cal N}=4$ supersymmetry.
There is only one dynamical scale in the problem: the distance or Higgs
expectation value, $M_{DYN}\sim u$.
Spontaneous supersymmetry breaking on the spectator branes is modeled
by
considering a thermal state (with temperature T).
The supersymmetry breaking scale is $M_{SUSY}\sim T$.
Standard supertrace formulae imply that when  ${\cal N}=4$
supersymmetry
is broken the vacuum energy scales as $M_{SUSY}^4$.
Here we find that the cosmological constant induced in our universe
due to supersymmetry breaking on the spectator brane is much smaller:
it scales as $\Lambda\sim M_{SUSY}^8/M_{DYN}^4$\footnote{For
$M_{SUSY}\sim 10
TeV$ and $M_{DYN}\sim M_{Planck}$ we obtain $\Lambda \sim 10^{-120}
M_{Planck}^4$.}.
Moreover, it is expected that the extra contributions to the vacuum
energy
on the universe brane due to loops of brane fields will have extra
suppression
factors of the coupling constant.
If instead we use a stack of black-branes whose extremal limit are 
branes at an orbifold singularity with $1\leq {\cal N}\leq 4$
then there is a similar behavior in the vaccum energy.
There is also a hint that this supression of the vacuum energy
would persist
for the case of ${\cal N}=2$ world-volume supersymmetry on the probe.

$\bullet$ \underline{Induction of field-dependent
 or time-varying speed of light on
a brane}.
When a probe brane moves in the gravitational field of another black
brane,
the induced field theory on the probe brane although Lorentz invariant,
has
a field dependent velocity of light.
In the gauge theory picture this effective velocity of light is due to
thermal
quantum effects.
In the simple example we analyze here this velocity of light dependents
on the
distance (which is also a dynamical scalar field of the probe brane) to
the
black-brane.
In the case discussed in this paper, the probe brane is outside the
horizon of
the black brane.
There may be situations where the effective velocity of light is time
dependent
and that can be achieved by taking the brane inside the horizon and
performing
the appropriate analytic continuation.
A time-varying speed of light can be an alternative to inflation and
can
thus provide different way to solve  the flatness problem in cosmology,
\cite{varc}.

The structure of this paper is as follows;
In section two we describe the basic black-brane solutions we will be
using here
as well as we evaluate the wold-volume probe action in such
backgrounds.
In section three we focus on three-branes, we take the near-horizon
limit and
gravitationally derive the thermal Yang-Mills potential.
In section four, we perform the one-loop computation of the potential
in
open string
theory and by taking the $\alpha'$-limit in thermal gauge-theory.
This agrees with the gravitational calculation up to exponentially
suppressed
terms.
In section five we do a similar analysis for two and four derivative
effective
couplings on the probe brane.
Finally in section six we discuss potential phenomenological
applications
of the phenomena discussed here.
In the appendix a careful evaluation of the RR gauge field in
black-brane
configurations is given.

\section{Black Dp-branes}
\setcounter{equation}{0}

We consider now the background geometry (in the string frame) of a
near-extremal black hole
describing a number of coinciding Dp-branes \cite{hs}:
\be
ds_{10}^2={-f(r)dt^2+d\vec x\cdot d\vec x\over
\sqrt{H_p(r)}}+\sqrt{H_p(r)}\left({dr^2\over
f(r)}+r^2d\Omega_{8-p}^2\right)\label{dmet}
\ee
where
\be
H_p(r)=1+{L^{7-p}\over r^{7-p}}\;\;\;,\;\;\;f(r)=1-{r_0^{7-p}\over
r^{7-p}}
\ee
The parameters $L$ and $r_0$ determine the AdS throat size
and the position of horizon, respectively. They are related
to the ADM mass $M$ and the (integer) Ramond-Ramond charge N
in the following way:
\be
M={\Omega_{8-p}V_p\over
2\kappa_{10}^2}~\left[(8-p)r_0^{7-p}+(7-p)L^{7-p}\right]\label{mp}
\ee
\be
N={(7-p)\Omega_{8-p}\over
2\kappa^2_{10}T_p}~L^{(7-p)/2}\sqrt{r_0^{7-p}+L^{7-p}},
\label{charge}\ee
where $\Omega_{n}$ is the volume of a unit $n$-dimensional sphere
$S^n$,
\be\Omega_{n}={2\pi^{(n+1)/2}\over\Gamma((n+1)/2)}\ , \ee
and $V_p$ is the common $p$-dimensional D-brane (flat) volume.
The  relations (\ref{mp},\ref{charge})
involve the D-brane tension $T_p$ and the 10-dimensional
gravitational constant $\kappa_{10}$ which are determined by
the string coupling $g_s$ and the string tension $\alpha'$ as follows:
\be
T_p={1\over (2\pi)^p \alpha'^{(p+1)/2}g_s}
\qquad,\qquad 2\kappa_{10}^2=(2\pi)^7\alpha'^4g_s^2\ .
\ee
The RR charge N is quantized, with each D-brane carrying a unit charge
so that N is equal to the number of D-branes.
Note that in the extremal case ($r_0=0$), $M=NV_pT_p$.
Finally,
\be
L^{7-p}=\sqrt{\left({2\kappa_{10}^2T_pN\over
(7-p)\Omega_{8-p}}\right)^2
+{1\over 4}r_0^{2(7-p)}}-{1\over 2}r_0^{7-p}
\ee

The RR charge is the source of the $p$-form field
\be
C_{012\cdots p}(r)~=~{2\kappa^2_{10}T_pN\over
\Omega_{8-p}(7-p)(r^{7-p}+L^{7-p})}~=~\sqrt{1+{r_0^{7-p}
\over L^{7-p}}}~{H_p(r)-1\over H_p(r)}\ .
\label{C}\ee
All other components vanish, except in the case of $p=3$,
when the self-duality condition
\be
{}F_{\mu_1\cdots\mu_5}={1\over
5!\sqrt{\det g}}
\epsilon_{\mu_1\cdots\mu_5\nu_1\cdots\nu_5}F^{\nu_1\cdots\nu_5}
\label{self}\ee
requires non-zero $p$-form components in the transverse
directions.
Since there are discrepancies in the literature,
we discuss the $p$-form solutions in more detail in the Appendix.
There is also a dilaton background (constant for $p=3$):
\be e^{\phi}=H^{(3-p)/4}_p(r) \label{ddil}\ee

By using standard methods of black hole thermodynamics,
it is straightforward to determine the
Hawking temperature,
chemical potential and entropy corresponding to the
solution (\ref{dmet},\ref{C},\ref{ddil}). They are respectively:
\be
T={7-p\over 4\pi}~{r_0^{(5-p)/2}\over
\sqrt{r_0^{7-p}+L^{7-p}}}\;\;\;,\;\;\;
\Phi=V_p T_p~{L^{(7-p)/2}\over \sqrt{r_0^{7-p}+L^{7-p}}}
\label{th}
\ee
\be
S={4\pi\Omega_{8-p}V_p\over
2\kappa_{10}^2}~r_0^{(9-p)/2}\sqrt{r_0^{7-p}+L^{7-p}}\ ,
\ee

We consider now a Dp-brane probing the above solution,
with zero background values for all other fields. In this case, the
D-brane
probe action is\footnote{There are also curvature depended
CP-odd
couplings.
These give zero contribution for the background at hand.}
\be
\Gamma_p=T_p~e^{-\phi}\int \sqrt{\det \hat g}+T_p\int \hat
C\label{gdp}
\ee
where we have also set the world-volume gauge field strength to zero,
$F_{\alpha\beta}=0$.
Using the metric (\ref{dmet}), the $p$-form (\ref{C}) and the
dilaton (\ref{ddil}), we obtain the static potential \cite{mal}
\be
V(r)=V_p T_p\left[{\sqrt{f(r)}\over H_p(r)}+C(r)\right]=V_p
T_p\left[{\sqrt{f(r)}\over H_p(r)}+\sqrt{1+{r_0^{7-p}\over
L^{7-p}}}{H_p(r)-1\over H_p(r)}\right]\label{4}\ee
where $C(r)\equiv C_{012\cdots p}(r)$.
The values of the potential at infinity
and at the horizon are, respectively,
\be
V(\infty)=V_pT_p\;\;\;,\;\;\;V(r_0)=\Phi\label{vs}
\ee

We  can expand the interaction potential $V^{\rm
int}(r)=V(r)-V(\infty)$
at large $r$, to obtain
\be
{V^{\rm int}(r)\over V_p
T_p}=\left[L^{7-p}\left(\sqrt{1+{r_0^{7-p}\over
L^{7-p}}}-1\right)-{1\over 2}r_0^{7-p}\right]{1\over r^{7-p}}+
\label{pot1}\ee
$$
-{1\over 8}\left[8L^{2(7-p)}\left(\sqrt{1+{r_0^{7-p}\over L^{7-p}}
}-1
-{1\over 2}{r_0^{7-p}\over L^{7-p}}\right)
r_0^{2(7-p)}\right]{1\over r^{2(7-p)}}+{\cal O}(r^{-3(7-p)})
$$

The leading long-distance term can be understood as follows:
it is due to the classical interaction of the extremal probe with the
non-extremal collection of p-branes. This interaction is proportional
\cite{db} to $Q_{\rm probe}(M_p-NV_pT_p)$.
An important point here is that the mass $M_p$, felt by the D$p$-brane
is
not the
same as the thermodynamic mass (\ref{mp}). From (\ref{pot1}) we obtain
\be
{M_p\over V_p}={(7-p)\Omega_{8-p}\over 2\kappa_{10}^2}\left[
L^{7-p}+{1\over 2}r_0^{7-p}\right]
\ee
and we have also $Q_{\rm probe}=V_pT_p$.
Thus,
\be
V^{\rm int}=-{2\kappa_{10}^2\over (7-p)\Omega_{8-p}}{Q_{\rm
probe}(M_p/V_p-NT_p)\over r^{7-p}}+\cdots
\ee
Different D-brane probes feel different masses.
The interaction potential for a D$(p-2n)$-brane probe
is
\be
V^{int}_{p,n}(r)=V_{p-2n}T_{p-2n}\left[		\sqrt{f(r)}H_p(r)^{-1+n/2}-1\right]=
\ee
$$
=-V_{p-2n}T_{p-2n}{\left(1-{n\over 2}\right)L^{7-p}+{1\over
2}r_0^{7-p}
\over r^{7-p}}+{\cal O}(r^{-2(7-p)})
$$
When $n\not =0$ there is no interaction due to the exchange of a RR
field
and the large distance interaction is
\be
V^{\rm int}_{p,n}(r)=-{2\kappa_{10}^2\over (7-p)\Omega_{8-p}}
{Q_{\rm probe}(M_{p,n}/V_p)\over r^{7-p}}+\cdots
\ee
with the mass seen by the D$(p-2n)$-branes
\be
{M_{p,n}\over V_p}={(7-p)\Omega_{8-p}\over 2\kappa_{10}^2}\left[
\left(1-{n\over 2}\right)L^{7-p}+{1\over 2}r_0^{7-p}\right]
\ee
Note that for a $(p-4)$-brane probe, $n=2$, the apparent mass vanishes
at
extremality, as expected, since the system of $p$- and $(p-4)$-branes
does
not break all of supersymmetry.
Also, note that for $n>2$, the interaction can become repulsive
near extremality
since for $r_0\ll L$ we obtain
that the leading large distance interaction  is
\be
V^{\rm int}_{p,n=0}(r)=-V_pT_p{r_0^{2(7-p)}\over 8L^{7-p}}{1\over
r^{7-p}}
\;\;\;,\;\;\;V^{\rm int}_{p,n}(r)=\left({n\over
2}-1\right){L^{7-p}V_{p-2n}T_{p-2n}\over r^{7-p}}
\ee
plus terms that are suppressed by extra powers of ${L^{7-p}\over
r^{7-p}}$,
${r_0^{7-p}\over L^{7-p}}$.
Finally, the potential for a fundamental string probe is similar to
that
of a $(p-2)$-brane.

\section{D3-branes, the near-horizon limit and the thermal Yang-Mills
potential}
\setcounter{equation}{0}

The case of D3-branes is particularly interesting because
the world-volume action of $N$ coinciding D-branes
involves a four-dimensional ${\cal N}{=}4$
supersymmetric $SU(N)$ Yang-Mills theory.
Moreover, in this case there is a natural correspondence (at all
scales) with
supergravity:
according to the Maldacena
conjecture
\cite{mald}, the large $N$ limit of this gauge theory is related
to the near-horizon AdS geometry of the extremal ($r_0=0$) black
D3-brane
solution (\ref{dmet}). Witten \cite{w2} has
exploited the AdS/SYM correspondence in order to study the large $N$
dynamics
of non-supersymmetric SYM, with ${\cal N}{=}4$ supersymmetries
broken by non-zero temperature effects. According to this proposal,
the non-extremal solution (\ref{dmet}) may be used to study
SYM at $T$ identified with the Hawking
temperature (\ref{th}) as long as $T\ll 1/L$, so that the metric
remains near-extremal $(r_0\ll L)$. In the near-horizon limit,
$\alpha'\equiv
l_s^2\to 0$ at $u\equiv r/\alpha'$ and $T$ fixed, the solution
(\ref{dmet})
describes an AdS-Schwarzschild black hole \cite{hp}:
\be
ds^2=l_s^2\bigg[{u^2\over R^2}(-f(u)dt^2+d\vec x\cdot d\vec x) +R^2
{du^2\over u^2f(u)}+R^2 d\Omega_{5}^2\bigg]+{\cal O}(l_s^4)\
,\label{met}
\ee
where
\be f(u)=1-{u_0^{4}\over u^{4}}\qquad ,\qquad
R^4\equiv 4\pi g_s N=\lambda\qquad,\qquad u_0=\pi T R^2\, ,\ee
where $\lambda$ is the 't Hooft coupling.
The limiting value of the four-form (\ref{C}) is
\be
C_{0123}=1+l_s^4\left({(\pi T R)^4\over 2}-{u^4\over R^4}\right)+
{\cal O}(l_s^8),
\label{Cc}\ee
The
four-form diverges at the boundary of $AdS_5$, $u\to \infty$.

A D3-brane probe in the bulk of the AdS space corresponding
to $N$ background D3-branes can be thought of as a realization
of $SU(N+1)$ gauge theory in the $SU(N)\times U(1)$ symmetric Higgs
phase.
In the following, we will examine the potential induced on the probe
brane
in the near-horizon limit and we will eventually compare it with the
gauge
theory calculation.

To that end, we will use the following expansions in the string length
scale
$l_s$:
\be
L^4=R^4 l_s^4\left(1-{1\over 2}\pi^4R^4T^4l_s^4\right)+{\cal
O}(l_s^{12})\ ,
\ee
\be
r_0=\pi T R^2 l_s^2\left(1+{1\over 4}\pi^4 T^4 R^4l_s^4+{\cal
O}(l_s^{8})\right)\ .
\ee
which follow from relations written in the previous section.

Taking the limit in the static potential (\ref{4}), we
obtain
\be
V(u)=V_3T_3\left\{1+l_s^4{u^4\over R^4}\left[\sqrt{1-\left({\pi T
R^2\over u}\right)^4}-1+{1\over 2}\left({\pi T R^2\over
u}\right)^4\right]+{\cal O}(l_s^8)\right\}
\ee
so that the interaction energy is
\be
V^{\rm int}(u)=V(u)-V_3T_3={V_3\over (2\pi)^3g_s}{u^4\over R^4}
\left[\sqrt{1-\left({\pi T
R^2\over u}\right)^4}-1+{1\over 2}\left({\pi T R^2\over
u}\right)^4\right]+{\cal O}(l_s^4)
\label{pot}\ee
and has a smooth limit as $l_s\to 0$.
Since the probe is BPS, $V^{\rm int}(\infty)=0$ \cite{1}.

The potential in (\ref{pot}) according to the Maldacena conjecture has
a direct
interpretation in the context of SU(N+1) N=4 gauge theory at finite
temperature.
We consider SU(N+1) N=4 gauge theory and a Higgs expectation value $u$
that
breaks the gauge symmetry to $SU(N)\times U(1)$.
At finite temperature, supersymmetry is broken. We consider now the
quantum
effective action for the U(1) factor, obtained by integrating out
all SU(N) as well as massive degrees of freedom.
This has an expansion in powers of $1/N$.
The leading piece is ${\cal O}(N)$.
At large 't Hooft coupling $\lambda$ this should be given by
supergravity
as in (\ref{pot}). We will rewrite it in terms of gauge theory
variables as
\be
V^{\rm int}(u)={NV_3\over 2\pi^2 }{u^4\over \lambda^2}
\left[\sqrt{1-{\pi^4\lambda^2 T^4
\over u^4}}-1+{1\over 2}{\pi^4\lambda^2 T^4 \over
u^4}\right]+{\cal O}(\lambda^{-3/2})
\label{pot22}\ee
This form should be compared with gauge theory for large Higgs
expectation
value $u>>T$.
For $u$ close to the horizon, there is a non-trivial map between the
supergravity and the gauge theory variable \cite{mal}.
This is obtained by matching the form of the world-volume kinetic terms
between supergravity and gauge theory.
The D-brane coordinate $\rho$ is related to the supergravity coordinate
$u$ as
\be
\rho^2=u^2+\sqrt{u^4-u_0^4}
\label{rn1}
\ee
\setcounter{footnote}{0}
Defining as in \cite{ty} the mass scale $M=(\rho-u_0)/R^2$
that controls the approach to the horizon we obtain\footnote{
This is in agreement with \cite{ty}.
The potential here differs however in the large u limit. This is because we chose the constant in the four-form (before taking the near-horizon
limit) to be such that it vanishes at infinity. A different prescription
was used in \cite{ty}. }
\be
V^{\rm int}=-{1\over 4}\pi^2NV_3T^4\left[1-{4\over \pi}{M\over T}-
{10\over \pi^2}
{M^2\over T^2}-{20\over \pi^3}{M^3\over T^3}+{\cal O}
\left({M^4\over T^4}\right)\right]
\label{sma}\ee

As mentioned before the potential in (\ref{pot22})
is supposed to be valid in the
limit of large 't Hooft coupling.
We can, however, consider the limit in which the Higgs expectation
is much larger than the thermal wavelength, $u>>\sqrt{\lambda} T$.
In this limit, supersymmetry is broken very softly.
Expanding the supergravity generated potential we obtain
\be
V^{\rm int}(u)={\pi^2\over 2}NV_3 T^4\sum_{m=1}^{\infty}
{(2m-1)!!\over 2^{m+1}(m+1)!}\left({\pi \lambda T^2\over
u^2}\right)^{2m}
\label{pot3}\ee
We observe that the large $u$ expansion is equivalent here to an
expansion
in the 't Hooft coupling.
The leading term $m=1$ is a three-loop term.
This is suggestive that although the result is strictly speaking valid
for
strong 't Hooft coupling, it might still be reliable also in
perturbation theory.

\begin{figure}
\begin{center}
\leavevmode
\epsfxsize=8cm
\epsffile{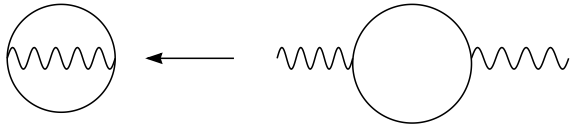}
\end{center}
\caption[]{\it ${\cal O}(N^2)$ two-loop contribution to the effective
potential.}
\label{f1}\end{figure}

\begin{figure}
\begin{center}
\leavevmode
\epsfxsize=5cm
\epsffile{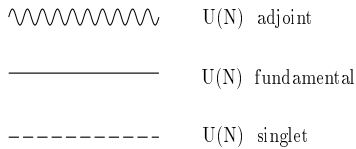}
\end{center}
\caption[]{\it Line notation for the Feynman diagrams.}
\label{f2}\end{figure}

This situation is reminiscent of an analogous phenomenon in the
near-extremal
 calculations of grey-body factors in D-brane black-holes \cite{gray}.
There also, a result that is valid at strong 't Hooft coupling agrees
with
Yang-Mills perturbation theory.
The reason was traced at specific non-renormalization theorems due to
softly
broken supersymmetry, \cite{nonr}.
A similar phenomenon is occurring here.
For this to work, contributions that are
higher order in $\alpha'$ should be suppressed by extra powers of
$\lambda
T^2/u^2$.
The next order term in the bulk action is of order $\alpha'^3$. To
calculate
its contribution to the potential we need the $\alpha'$ corrected form
of the
four-form away from the near-horizon limit. Such an investigation is
under way
and will provide further evidence for the above.

The diagrams relevant here, that are leading in $N$,
at higher orders in the 't Hooft
coupling
have one boundary stuck at
the probe brane and the rest are attached to the stack of $N$
D3-branes.
Thus, they scale as $N(g_sN)^{B-2}\sim N\l^{B-2}$ where $B$ is the
total
number
of boundaries.
Moreover, the supergravity result implies that only diagrams with
$B=2m+2$,
$m>0$ contribute.
In the gauge theory this implies that contributing diagrams appear at
order $m+1$,
starting at three loops. Written in the double line notation of 't
Hooft
they have $2m+1$ index loops with $N$ circulating colors and one with
one
color.
Thus, they can be represented as a disk with $2m+1$ holes.
Cutting the diagrams appropriately we can see that $SU(N)$ degrees of
freedom
circulate in intermediate channels. These are responsible for
producing
the temperature dependent factors and in particular the power-like
behaviour of
the potential.

In analogy with grey-body factor calculations we can conjecture
that the potential (\ref{pot22}) can be compared with perturbative
thermal
Yang-Mills calculations.
In particular this implies that there are no one-loop and two-loop
contributions, and the first non-trivial term is coming in at
three-loops.
The leading contribution is $\sim \lambda^2~T^8/u^4$.

In the next section we will do the one-loop computation and show that
in the
large $u$ limit it is exponentially suppressed.
These exponential contributions are "non-perturbative" from our point
of view
and can be interpreted as due to solitonic loops.
Thus, up to such terms, the gauge theory will give a zero contribution
at one
loop.
Moreover, there is independent evidence that the next term comes from
three
loops and behaves like $T^8/u^4$.
In \cite{wir} the theory analyzed is SU(2) ${\cal N}=2$ super YM and
the relevant contribution was found from a two-loop diagram with an
effective one-loop generated vertex.
This amounts to a three-loop contribution in the original SU(2) theory.

There is an alternative way to implement the perturbative calculation:
Consider a thermal ensemble in the SU(N) part but treat the massive
degrees
of freedom as though they were not thermal.
Then the one loop calculation in the gauge theory will give zero
both for the potential and the kinetic terms. This will be shown
explicitly in the next section.
The difference with the previous calculation is that there will be no
exponentially suppressed terms here.

The above can produce a stringent test of the Maldacena conjecture that
can
be
performed in Yang-Mills perturbation theory.
One needs to evaluate the effective potential of the $U(1)$ scalar in
$SU(N+1)\to SU(N)\times U(1)$ theory at two and three loops.
The relevant ${\cal O}(N^2)$ two-loop diagram
is shown in fig. \ref{f1} (we use the notation of fig. \ref{f2}).
This sumarises contributions of scalars, fermions
and ghosts. Note that there is no relevant double-bubble diagram.
According to our conjecture this should give only exponentially small
contributions at large $u$, (or zero iff the massive fundamental is not
thermalized).
The diagram includes two massive external propagators as well as a
massless
internal one. This can be written as a one-loop diagram of the
SU(N) degrees of freedom where instead of the tree propagator we use
the one-loop corrected propagator where only massive states go around
the loop (fig. \ref{f1}).
However, the results of the next section imply that such corrections to
the
propagators after summed over the various states are exponentially
suppressed.
This is in accord with our expectations that the two-loop contribution
to the
potential is exponentially suppressed.

\begin{figure}
\begin{center}
\leavevmode
\epsfxsize=8cm
\epsffile{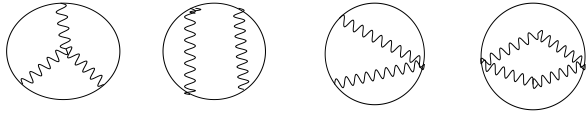}
\end{center}
\caption[]{\it ${\cal O}(N^3)$ three-loop contributions to the
effective potential.}
\label{f3}\end{figure}

The appropriate field theory diagrams that contribute to three-loop
order 
are shown in fig. \ref{f3}.
It would be extremely interesting to perform the field theory
calculation and reproduce the leading term in (\ref{pot3}).

The conjecture above is also valid in the case where $SU(N+M)\to
SU(N)\times
SU(M)\times U(1)$ with $N>>M$.
The probe action must be replaced by the appropriate
non-abelian generalization \cite{tsen}
and the potential is $M^2$ times the one calculated in the U(1) case.

\section{Perturbation theory and the potential}
\def\l{\lambda}
\setcounter{equation}{0}

In this section we will study the static potential further and we will
try to
make contact with  gauge theory perturbation theory.

We will first consider an open string theory evaluation of the
static potential.
This can be done by integrating out in perturbation theory
the stretched strings between $N$ D3-branes and a
D3-probe at distance $r$.
The one-loop free-energy at finite temperature is \cite{tse}
\be
F(\beta/l_s,r/l_s)=-{V_3 N\over
16\pi^2\alpha'^2}\int_{0}^{\infty}{dt\over t^3}e^{-{r^2t\over
2\pi\alpha'}}\left[\vartheta_3\left({i\beta^2\over 4\alpha' \pi
t}\right){1\over 2}
{\vartheta^4_{3}(it)-\vartheta^4_4(it)\over \eta^{12}(it)}-
\right.
\ee
$$-\left.\vartheta_4\left({i\beta^2\over 4\alpha' \pi t}\right){1\over
2}{\vartheta_2^4(it)\over \eta^{12}(it)}\right]
$$
Using supersymmetry and
\be
\vartheta_3\left({i\beta^2\over 4\alpha' \pi
t}\right)-\vartheta_4\left({i\beta^2\over 4 \alpha'\pi
t}\right)=2\vartheta_2\left({i\beta^2\over \alpha' \pi t}\right)
\ee
we obtain
\be
F=-{V_3 N\over 16\pi^2\alpha'^2}\int_{0}^{\infty}{dt\over
t^3}e^{-{r^2t\over 2\pi\alpha'}}\vartheta_2\left({i\beta^2\over
\alpha'\pi t}\right)f_{\rm open-string}(t)
\label{1lpot}\ee
with
\be
f_{\rm open-string}(t)={1\over 2}{\vartheta_2^4(it)\over
\eta^{12}(it)}=8+{\cal
O}(e^{-\pi t})
\ee
\be
f_{\rm open-string}\left(1/t\right)=t^{-4}f_{closed-string}(t)=
{t^{-4}\over 2}{\vartheta_4^4(it)\over \eta^{12}(it)}={t^{-4}\over 2}
\left(e^{\pi
t}+{\cal O}(e^{-\pi t})\right)
\ee
Near $t\to\infty$ the integrand behaves as
\be
{1\over \beta}\int^{\infty} {dt\over t^{5/2}} e^{-{r^2t\over
2\pi^2\alpha'}}
\ee
which is convergent. If we consider the expansion of the
exponential
in powers of $r^2$ then the terms after the first two diverge.
Near $t\to 0$ the integrand behaves as
\be
 \int_0 tdt e^{{\pi\over t}\left(1-{\beta^2\over 4\alpha'\pi}\right)}
\ee
and we clearly see the signal of the open string Hagedorn transition
at
$\beta_H=2\sqrt{\pi\alpha'}$.
We will thus assume from now on that $\beta>\beta_H$.

Among the three scales $r,\beta, \alpha'$ we can make two
dimensionless
parameters, $\tilde\beta=\beta/\sqrt{\alpha'}$ and $\tilde
r=r/\sqrt{\alpha'}$.
We would like to find the behavior of $F$ as a function of
$\tilde \beta$ and $\tilde r$.

When $\tilde\beta \tilde r >>1$ the integral can be evaluated by
saddle
point.
The position of the saddle point is at $t_0={\beta\over \sqrt{2} r}$.
For $t_0>1$, we use $f_{\rm open-string}$ while for $t_0<1$
we use
$f_{\rm closed-string}$.
We obtain:

$\bullet$ $\tilde \beta \tilde r >>1$ and $\beta>r$
\be
F={V_3 N\over (2\pi)^{1/4}2\pi^2}\sqrt{r^3\over \beta^5\alpha'^{3}}
e^{-{\beta r\over \sqrt{2\pi}~\alpha'}}f_{\rm
open-string}\left(i\sqrt{\pi\over 2}{\beta\over r}\right)
\ee

$\bullet$ $\tilde \beta \tilde r >>1$ and $\beta<r$
\be
F={V_3 N\over 8(2\pi)^{1/4}}\sqrt{{\beta^3\over r^5\alpha'^{3}}}
e^{-{\beta r\over \sqrt{2\pi}~\alpha'}}f_{\rm
closed-string}\left(i\sqrt{2\over \pi}{r\over \beta}\right)
\ee
These are instanton contributions. The instanton is the world-sheet of
an open string stretched a distance $r$ and wound around the temporal
circle.
This gives a configuration with area $\beta r$ and action $S_{\rm
inst}\sim
\beta r/\alpha'$.
The determinant factor is in fact the determinant on the
above-mentioned
cylinder with (real) modulus $t\sim \beta/r$ of the open string
fluctuations.
When $t>1$ they are best described by open string states while when
$t<1$
they are best described by closed string states.

Since $\beta>\beta_H\to \tilde\beta>2\sqrt{\pi}$, the only other
corner
in the
$\tilde \beta$,$\tilde r$ plane to investigate is $\tilde\beta\tilde
r<1$.
In the small $r$ region the massive modes of the string are
subleading.
The reason is that as $r\to 0$ the mass gap vanishes for the massless
states and it is those that dominate the behavior of the free energy.
The dominant contribution is \cite{the,ty}
\be
F= V_3 (2N)\left[-{\pi^2\over 6\beta^4}+{r^2\over 4\pi
\beta^2\alpha'^2}-{r^3\over
3\pi^2\sqrt{2\pi}\beta\alpha'^3}+\cdots\right]
\ee
which describes accurately the behavior of the free energy in the
region
$\tilde \beta\gg 1$, $\tilde\beta\tilde r\ll 1$.
This is the region with a temperature much lower than the string scale
and  a Higgs expectation value much lower than the temperature when
both are measured in string units.
The behavior is indeed that of field theory. The leading term is the
change of the Yang Mills free energy $\pi^2V_2 N^2 T^4$ as $N\to N+1$
and the subleading term generates a mass for the $u$-scalar,
$u=r/\alpha'$,
$m_u\sim g_sT$.
Moreover, the potential will force  $r$ to relax back to $r=0$.

\begin{figure}
\begin{center}
\leavevmode
\epsfxsize=8cm
\epsffile{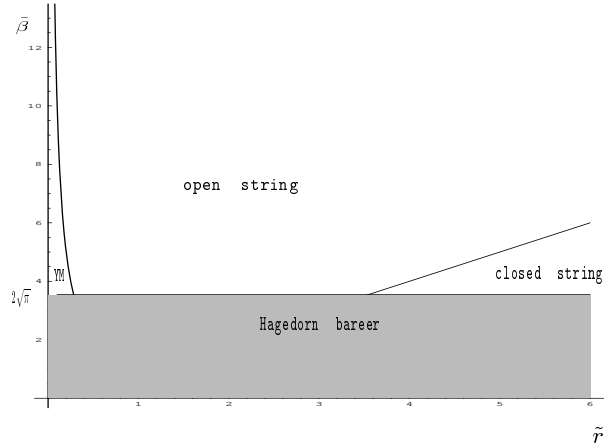}
\end{center}
\caption[]{\it The various regions of the one-loop effective
potential}
\label{f4}\end{figure}

We will now go to the near-horizon limit: $u=r/\alpha'$ and $\beta$
fixed.
\be
F_M=-{V_3 N\over 16\pi^2\alpha'^2}\int_{0}^{\infty}{dt\over
t^3}e^{-{u^2\alpha't\over 2\pi}}\vartheta_2\left({i\beta^2\over
\alpha'\pi t}\right)f_{\rm open-string}(t)
\ee
After changing the variable $t\to t/\alpha'$,
\be
F_M=-{V_3 N\over 16\pi^2}\int_{0}^{\infty}{dt\over t^3}e^{-{u^2t\over
2\pi}}\vartheta_2\left({i\beta^2\over \pi t}\right)f_{\rm
open-string}(t/\alpha')=
\ee
$$
-{V_3 N\over \pi^2}\int_{0}^{\infty}{dt\over t^3}e^{-{u^2t\over
2\pi}}\vartheta_2\left({i\beta^2\over \pi t}\right)+{\cal O}(\alpha')=
-{V_3 N\over \beta^4}\int_{0}^{\infty}{dt\over
t^3}e^{-{u^2\beta^2t\over 2\pi^2}}\vartheta_2\left({i\over
t}\right)+{\cal O}(\alpha')
$$
Note that the Hagedorn boundary has been pushed to $T\to \infty$.
There are essentially two distinct regions:

$\bullet$ $\beta u\gg 1$. It corresponds to the open solitonic string
region
with
\be
F={8V_3 N\over (2\pi)^{1/4}\pi^2}\sqrt{u^3\over \beta^5}
e^{-{\beta u\over \sqrt{2\pi}}}
\ee
Here the Higgs expectation value is much larger than the temperature.
Thus, integrating out these massive modes gives exponentially
suppressed
contributions.
The supersymmetry breaking parameter,
namely the temperature, is much smaller than the scale $u$ of the
${\cal
N}{=}4$
supersymmetric theory. Thus, the supersymmetric non-renormalization
theorem for the potential
is true to exponential accuracy.

$\bullet$ $\beta u\ll 1$.
$$
F=-{V_3(2N)\pi^2\over 6\beta^4}\left[1-{3u^2\beta^2\over 2\pi^3}+
{u^3\beta^3\over \sqrt{2\pi}\pi^4}-{15\log 2\over 4}{u^4\beta^4\over
\pi^6}
-\right.
$$
$$-
\left.{u^6\beta^6\over \pi^9}\sum_{n=0}^{\infty}\left({u^2\beta^2\over
2\pi^2}
\right)^n\left({{3\over 2}+n+2\atop
n+3}\right)(1-2^{-(2n+2)})\zeta(2n+3)\right]
$$
\be
=V_3 (2N)\left[-{\pi^2\over 6\beta^4}+{u^2\over 4\pi
\beta^2}-{u^3\over 3\pi^2\sqrt{2\pi}\beta}\cdots \right]
\label{sma1}\ee
Here the Higgs expectation value is much smaller than the temperature.
The theory is close to unbroken Yang Mills as indicated by the free
energy.
Supersymmetry is completely broken and this is reflected in the form
of the
free energy.
The minimum of the potential is again at $u=0$.
This is true for any temperature since the derivative of the potential
with respect to $u$ is proportional to $u$ times an integral of a
non-negative
integrand.
 The only other extremum  is at $u=\infty$ but this is a maximum.
Thus, the potential is monotonically decreasing from $u=\infty$ to
$u=0$.

The form of the potential for large temperature (\ref{sma1}) agrees
qualitatively with the one obtained by supergravity. In this region we
do not expect quantitative agreement, but it is obvious that the
one-loop
contributions has similar features to the strong coupling
(supergravity)
result.
In this limit the vacuum energy has a leading piece which is the
supersymmetry
breaking scale (temperature) to the fourth power, as expected in  a
non-supersymmetric theory.

The situation is different at large distances, or large Higgs
expectation values.
The one-loop results is exponentially suppressed, and one reason for
this is
approximate supersymmetry in this limit.
This can be clearly seen by our computation. Separately bosons and
fermions
produce a temperature-independent polynomial piece scaling as $r^4$ at
large $r$ plus exponentially
small contributions. The polynomial piece cancels between fermions and
bosons and
we are left with the exponentially small contribution.
In general this is an expected feature of softly or spontaneously
broken
supersymmetry.
A different way of viewing this behavior is to state that the object
(open string) going around the
loop  has a finite size and thus gives an exponentially small
contribution.
{}From the point of view of supergravity it behaves as a soliton, and
it is not
visible in the long distance expansion.

It is obvious from our calculation that if the massive open string
(or in the $\alpha'\to 0$ limit the massive fundamental multiplet)
is at zero temperature then the contribution to the potential will be zero.

The potential (vacuum energy) is generically expected to scale as the
larger
mass in the fourth power, namely $r^4$.
{}From supergravity we obtain the prediction that the leading
contribution
is $T^8/r^4$ which is strongly
suppressed.

We conclude that the supergravity calculation provides us with the
physical
(non-Wilsonian) effective action of the $U(1)$ gauge theory after
integrating
out the $SU(N)$ plus the massive degrees of freedom.
Moreover, exponentially small contributions in perturbation theory are
not visible in the supergravity result.
This suggests that the proper context is to take only the SU(N) part to
be in a thermal bath.
The probe brane is certainly at zero temperature and our analysis indicates
that at large distance the ultra-massive string can also be taken
to be at zero temperature.
  
The quantum contributions of
the world-volume fields have not been taken into account.
In the large $N$ limit
they turn out to be subleading.

\setcounter{footnote}{0}

\section{Kinetic and Quartic terms}
\setcounter{equation}{0}

We will now expand the D3-brane action keeping also (bosonic)
quadratic  terms in
derivatives.
We obtain
\be
S_{D3}=T_3 \int \sqrt{-\det (\hat
g+(2\pi\alpha')F-B)}+T_3\int \hat
C=S_0+S^s_{1}+S^F_{1}+{\cal O}(v^4)
\label{ac}\ee
with $S_0$ being the potential that we already discussed. $S^s_1$
represents
the
kinetic terms for the scalars,
\be
S^s_1={T_3\over 2}\int d^{4}x \left[{1\over f(r)}\partial r\cdot
\partial
r+r^2h_{\alpha\beta}\partial y^{\alpha}\cdot \partial y^{\beta}\right]
\ee
where
\be
\partial \phi_1\cdot \partial\phi_2=\tilde
g^{\mu\nu}\partial_{\mu}\phi\partial_{\nu}\phi\equiv -{1\over
\sqrt{f(r)}}\partial_0 \phi_1\cdot
\partial_0\phi_2+\sqrt{f(r)}\sum_{i=1}^3\partial_i \phi_1\cdot
\partial_i\phi_2
\ee
and $S^F_1$ are the gauge kinetic terms,
\be
S_F^1=(\pi\alpha')^2T_3\int d^{3}x {1\over \sqrt{f(r)}}\left[\tilde
g^{\mu\rho}\tilde g^{\nu\sigma}F_{\mu\nu}F_{\rho\sigma}\right]=
2(\pi\alpha')^2T_3\int d^{3}x \left[{1\over \sqrt{f(r)}}\vec
E^2+\sqrt{f(r)}\vec B^2\right]
\ee
where as usual $E_i=F_{0i}$, $B_i=\epsilon_{ijk}F_{jk}/2$.
We have also set $B_{\mu\nu}=0$.
Note that the velocity of light on the 3-brane is $r$-dependent.
{}From the formulae above we can ascertain that
\be
c_{\rm eff}=\sqrt{f(r)}=\sqrt{1-\left({r_0\over r}\right)^4}
\label{vl}\ee
For a brane next to the horizon, any time dependent
fluctuation
freezes, since the velocity of light vanishes there.

To see this, set $\epsilon=r-r_0\to 0$ and
\be
L_{kin}(r)\sim -{\dot r^2\over f^{3/2}}+{(\nabla r)^2\over f^{1/2}}\to
-\left(4{r_0\over \epsilon}\right)^{3/2}\dot
\epsilon^2+\left(4{r_0\over \epsilon}\right)^{1/2}(\nabla \epsilon)^2
=
\ee
$$
={32 \sqrt{r_0}\over 9}\left(-4r_0{\dot z^2\over z^{4/3}}+(\nabla z)^2
\right)
$$
with $\epsilon=z^{4/3}$.
A similar thing happens for the $S^5$ coordinates.
For the gauge fields
\be
L_F\sim \sqrt{r_0\over \epsilon}\vec E^2+ \sqrt{\epsilon\over r_0}\vec
B^2
\ee
and near the horizon $\vec E\to 0$. Since $\vec E$ is the source for
fundamental strings attached to the brane, that implies that such
couplings are
suppressed close to the horizon.
Taking the near horizon limit does not essentially modify the kinetic
terms we
have described above.

The effective kinetic terms can be examined at large u in the same way
done for
the potential.
The corrections are power series in $(TR^2/u)^4$ and they start at two
loops.
An open string calculation similar to that for the potential can be
done
also for the kinetic terms along the lines of \cite{bach,tse}.
The result for the one-loop correction to the two-derivative terms is
\be
F_2\sim -{N\over 8\cdot 16\pi^2}\int_{0}^{\infty}{dt\over
t}e^{-{r^2t\over 2\pi\alpha'}}\vartheta_2\left({i\beta^2\over
\alpha'\pi t}\right)f_{\rm open-string}(t)~g_2(t)
\label{1lkin}\ee
with
\be
g_2(t)=3{\vartheta''_2(it)\over \vartheta_2(it)}-4
{\vartheta'''_1(it)\over \vartheta'_1(it)}=2\pi^2+{\cal O}(e^{-2\pi
t})
\ee
In the near-horizon limit, this becomes
\be
F_2^M=-{N\over 64}\int_{0}^{\infty}{dt\over
t}e^{-{u^2\beta^2t\over 2\pi^2}}\vartheta_2\left({i\over
 t}\right)=
\ee
$$=-{N\over 64}(2\pi)^{3/4}(u\beta)^{-1/4}e^{-{u\beta\over
\sqrt{2\pi}}}
+\cdots
$$
where we assumed in the second line $u\beta\gg 1$.
Thus, at one-loop  the conclusion is that the correction to the
kinetic terms is
exponentially suppressed for large $u$ (as for the potential).
This is again a signal of softly broken supersymmetry.
Also, the insertion of the two vertex operators in the diagram reduces
the
effective supersymmetry to ${\cal N}{=}2$
 so that already at two loops there is a
power correction.
It is also obvious from our expression that if the massive string
is not thermal then the one-loop contribution to the kinetic terms
vanishes.

If one continues to the $F^4$ couplings, then there is already a
one-loop
contribution in the extremal limit, and it is corrected by a power
series
in  $(TR^2/u)^4 $ coming from higher loops.
The D-brane action (\ref{ac}) gives
\be
S_{F^4}=-(2\pi\alpha')^4T_3\int {H(r)\over
8f(r)^{3/2}}\left[(F^2)^2-F^4\right]
\ee
where
\be
F^4=F_{\mu\nu}{F^{\mu}}_{\rho}{F^{\rho}}_{\sigma}
F^{\nu\sigma}\;\;\;,\;\;\;
F^2={1\over 2}F_{\mu\nu}F^{\mu\nu}
\ee
and the contractions above are made with the effective metric in
$\tilde g$.
We have
\be
F^2=\vec B^2-{1\over f}\vec E^2\;\;\;,\;\;\;(F^2)^2-{1\over
2}F^4=-{2\over f}
(\vec E\cdot\vec B)^2
\ee
In the near-horizon limit, it becomes
\be
S_{F^4}={2\pi^2 N}\int d^4x {(\vec E\cdot\vec B)^2\over
u^4\left(1-{u_0^4\over u^4}\right)^{5/2}}={2\pi^2 N}\int d^4x {(\vec
E\cdot\vec B)^2\over
u^4}+{\cal O}(T^4 R^8)
\ee
The leading term is the one-loop contribution while the next term
appears at
three loops.

The one-loop D-brane contribution to the $F^4$ coupling is
\be
F_4= {N\alpha'^2\over 2}\int_{0}^{\infty}dt~t~e^{-{r^2t\over
2\pi\alpha'}}\left[\vartheta_3\left({i\beta^2\over
\alpha'\pi t}\right)+\vartheta_2\left({i\beta^2\over
\alpha'\pi t}\right)f_{\rm open-string}(t)~g_4(t)\right]
\label{11lkin}\ee
with
\be
g_4(t)={1\over \pi^4}\left[15{\vartheta''''_2(it)\over
\vartheta_2(it)}
+32\left({\vartheta'''_1(it)\over \vartheta'_1(it)}\right)^2
-48{\vartheta^{(5)}_1(it)\over \vartheta'_1(it)}-120{\vartheta''_2(it)
\over \vartheta_2(it)}
{\vartheta'''_1(it)\over \vartheta'_1(it)}\right]=-121+
{\cal O}(e^{-2\pi t})
\ee
becoming in the near-horizon limit
\be
F_4^M={N\beta^4\over 2\pi^2}\int_0^{\infty}dt~ t~
e^{-{u^2\beta^2t\over 2\pi^2}}\left[\vartheta_3\left({i\over
 t}\right)-8\cdot 121\vartheta_2\left({i\over
 t}\right)\right]={2\pi^2 N\over u^4}+
 {\cal O}(e^{-{u\beta\over \sqrt{2\pi}}})
\ee
which agrees with the leading order in the supergravity calculation.
We can conclude that all effects that vanish in the limit of exact
supersymmetry are exponentially suppressed in the presence  of
non-zero
temperature.

We can use the supergravity calculation to study the
effective electric and
magnetic couplings of the $U(1)$ obtained after the breaking of
$SU(N+1)\to
SU(N)\times U(1)$\footnote{As mentioned earlier
our results are valid for the
breaking
$SU(N+M)\to SU(N)\times SU(M)\times U(1)$ with $M\ll N$.}
In fact, since there is a potential for $u$, we cannot make this
discussion when the consider the $U(1)$ quantum theory.
So, we consider integrating out the $SU(N)$ degrees of freedom, and
treat
the $U(1)$ sub-theory as a source theory that is renormalized.
In the near-horizon limit\footnote{We assume imaginary time.}
$S_1^F$ implies the
following electric and magnetic effective couplings
\be
{1\over g^2_e(T,u)}={1\over 2\pi g_s f(u)}
\;\;\;,\;\;\;
{1\over g^2_m(T,u)}={f(u)\over 2\pi g_s}
\ee
with $f(u)=\sqrt{1-\left({\pi T\l^{1/2}\over u}\right)^4}$,
and $\lambda=2Ng^2_{YM}$ is the 't Hooft coupling.
The equations above can be rewritten in terms of the associated
't Hooft couplings
as
\be
\l_e(T,u)=\l f(u)\;\;\;,\;\;\;\l_m(T,u)={\l\over f(u)}\;\;\;,\;\;\;
\l_e(T,u)\l_m(T,u)=\l^2
\label{cou}\ee
The electric field component $E_i$ is the source for fundamental
strings
in the $x^i$ direction. The magnetic field $B_i$ is the source
for D-strings in the $x_i$ direction. S duality of the IIB theory,
interchanges F and D strings and explains why $\l_m/\l=\l/\l_e$.

We can view here the Higgs expectation value as the running scale and
the temperature as 
the cutoff (equivalently as the $\Lambda$ scale).\footnote{From 
the gauge theory point of view, the correct scale
is $\rho$ in (\ref{rn1}).
However, using $u$ does not change the qualitative behavior of the
effective couplings.}
The effective couplings satisfy the renormalization group equations
\be
u{\partial \l_e\over \partial u}=\beta_e(\lambda_e)=-2\lambda_e+
{2\lambda_m}\;\;\;,\;\;\;u{\partial \l_m\over
\partial u}=\beta_m(\lambda_m)=2\lambda_m-
2{\lambda^2_m\over \lambda_e}
\label{beta}\ee
with the ultraviolet coupling $\l_e(u=\infty)=\l_m(u=\infty)=\l$.
As one flows to the infrared, the electric coupling decreases while
the
magnetic coupling decreases.
When $u=\pi T\lambda^{1/2}$ (the horizon in supergravity)
the electric coupling
vanishes while the magnetic coupling blows up.
Thus, the electric coupling is IR free while the magnetic
coupling is IR strong.
The region inside the horizon describes the theory for energies below
the thermal cutoff. The region far from the horizon corresponds to
scales were supersymmetry is a
good approximation.

We can analyze the system of $\beta$-functions (\ref{beta}) beyond our
specific solution.
The general solution can be written in the form
\be
\l_e^2=C_1+{C_2\over u^4}\;\;\;,\;\;\;\l_m^2={C_1^2\over C_1+{C_2\over
u^4}}
\ee
Note that $C_1\geq 0$ if we require a reasonable
UV limit.
There are different types of behavior:

$\bullet$ $C_1=0$, then $C_2>0$ and $\l_m=0$ and $\l_e$ is an
asymptotically free coupling that blows up in the IR.

$\bullet$ $C_1>0$ and $C_2<0$. This behavior is the one realized in
the
D-brane system
with $\lambda^2=C_1$ and $C_2=-\pi^4T^4\l^2$.

$\bullet$ $C_1,C_2=0$. This the generic fixed point of this system of
$\beta$-functions with $\l_e=\l_m=$ constant.

$\bullet$ $C_1>0$, $C_2>0$. This situation is qualitatively similar to
the
$C_1=0$ case but the electric coupling has a non-zero UV value, while
the magnetic coupling is IR free.

The running effective couplings in (\ref{cou}) are the physical
effective
$U(1)$
couplings (as we argued in the previous section).
They constitute concrete predictions of supergravity for spontaneously
broken
Yang-Mills theory in the large $N$ limit.
It would be interesting if they could be compared to a different
approach.

\def\g{\gamma}
\def\mpl{M_{\rm P}}
\def\ms{M_{\rm s}}
\def\ls{l_{\rm s}}
\def\l{\lambda}
\def\gs{g_{\rm s}}
\section{Phenomenological Implications}
\setcounter{equation}{0}
\def\ll{\Lambda}

There are two effects presented above that merit an extra discussion.
The first is the fact that the leading behavior of the vacuum energy
induced in the probe brane scales as $T^8/u^4$.
To decipher this dependence we will have to remind the reader the
behavior of the vacuum energy in a spontaneously broken supersymmetric
theory.
The vacuum one-loop diagram (in four dimensions)
has a well-known dependence on the cutoff as well as the particle mass
\be
\int {d^4k\over \pi^2}\log(k^2+m^2)={\ll^4\over 2}
\left(\log \ll^2-{1\over 2}\right)+m^2\ll^2+{m^4\over
2}\left(\log{m^2\over
\ll^2}-{1\over 2}\right)-{1\over 3}{m^6\over \ll^2}+{1\over 8}{m^8\over
\ll^4}
+{\cal O}(\ll^{-6})
\ee
In a supersymmetric theory the leading $\ll^4$ terms are proportional to
$Str[1]=\#$ fermions-$\#$ bosons.
The subleading $\ll^2$ terms are proportional to $Str[m^2]$.
In a spontaneously broken global  theory the soft masses are 
proportional to
the helicities so that mass supertraces turn into 
helicity supertraces
\cite{fer,book}.
For an ${\cal N}=4$ theory $Str[m^2]=0$ and the leading term is
proportional
to $M_{SUSY}^4$.
For a ${\cal N}=8$ theory (supergravity)  with a cutoff, softly 
broken so that 
masses are again proportional to helicities
$Str[m^4]=Str[m^6]=0$
but the logarithmic divergence prohibits a further suppression of the
vacuum energy.
For Scherk-Schwarz \cite{ss} supersymmetry breaking in string theory \cite{ss2} as well as that induced 
by temperature (in string theory this is similar to the previous mechanism)
the vaccum energy for spontaneously broken ${\cal N}\geq 4$ supersymmetry
(in four dimensions) 
scales as $M_{SUSY}^4$ \cite{A}.

There has been a revival lately of the the idea  \cite{brane} that our
four dimensional universe is a
three-brane,
embedded in ten dimensional, (partially) compactified spacetime.
This was viable in situations with a very low string scale
\cite{A}-\cite{Ver}.
Moreover, other three-branes may be providing mirror universes.
A prototype of this  is the Ho\v rava-Witten interpretation of the
Heterotic String \cite{hw}.
A popular mechanism of supersymmetry breaking in our universe is that
supersymmetry breaks spontaneously (probably due to strong dynamics)
on a spectator brane and then this supersymmetry breaking is
communicated via gravity to the universe brane \cite{hor}.

Here we have a toy model of this situation.
The spectator brane is the black-brane
while our universe is represented by the probe brane.
Supersymmetry is broken in the far-away brane by thermal effects.
Although this might not be the exact way we would like
supersymmetry to be broken,
it does represents soft supersymmetry breaking.
Both three-branes carry ${\cal N}=4$ supersymmetry.
There is only one dynamical scale in the problem (when supersymmetry is
unbroken): the distance or Higgs
expectation value, $M_{\rm DYN}\sim u$.
The supersymmetry breaking scale is given by the temperature,
$M_{\rm SUSY}\sim T$.

When $ M_{\rm SUSY}\sim M_{\rm DYN}$ then supersymmetry is strongly
broken and
we have obtained for the vacuum energy the result (\ref{sma}) which
scales as
$M^4_{\rm SUSY}$. This is the natural expectation, as argued above,
from
broken ${\cal N}=4$ supersymmetry.
In the opposite limit where the supersymmetry breaking scale is much
smaller than the dynamical scale,  we find that the cosmological
constant induced in our universe
due to supersymmetry breaking on the spectator brane is much smaller:
it scales as $\sim M_{SUSY}^8/M_{DYN}^4$.
This  scaling is subleading to that expected in a finite theory with 
maximal supersymmetry as argued previously.

One comment is in order here.
We consider the probe brane to be initially at zero temperature (unbroken supersymmetry). The statement that the vaccum nergy induced on the brane
is supressed is equivalent to the statement that the probe brane does not thermalize at any given finite time. In fact we do expect this to be true,
since we are in the limit of large distance. Since the interactions of 
the probe brane with the thermal pile of branes are of gravitational
stregth we do not expect thermalization at finite times.
Note that this intuition implies that quantum corrections on the probe brane
will also be supressed.

In order to describe a concrete situation 
we  assume that our universe is a 
Dp-brane a distance $r$ away from
the black Dp-brane.
We assume that the (9-p) transverse dimensions are compactified with volume 
$V_{t}$ while the extra $p-3$ directions on the brane are compactified with 
volume $V_{||}$.
Thus, the four-dimensional Planck scale and low energy gauge coupling  are 
given by 
\be
\mpl^2={V_{t}V_{||} \ms^8\over \gs^2} \;\;\;,\;\;\;{1\over g_{YM}^2}=
{V_{||}\ms^{p-3}\over
\gs}
\label{def}\ee

For the solutions of the previous section to be applicable here we must
have  that the distance between the branes is much smaller than the linear
dimensions of the compact tranverse space:
\be
{r\over l_s}<<{V_t^{1/(9-p)}\over l_s}=\left[\left({\mpl\over \ms}\right)^{2}
 ~\gs g_{YM}^2\right]^{1/(9-p)}
\label{con}\ee
The effective induced cosmological constant on the universe-brane 
is given by
\be
{\Lambda_4\over \mpl^4}={V^{int}\over \mpl^4}
\label{cosmo1}\ee
where $V^{int}$ is given in  (\ref{pot1}).
In the cases we will be interested  we can approximate\footnote{We neglet
factors of two and $\pi 's$.}
\be
\xi_4\equiv{\Lambda_4\over \mpl^4}\sim {V_{||}\ms^{p+1}\over \gs\mpl^4}\left({r_0^2\over rL}\right)^{7-p}
\ee
For $p<5$ we find
\be
\xi_4\sim {V_{||}\ms^{p+1}\over \gs\mpl^4}\l^{{2(9-p)\over (5-p)}}
{(T~\ls)^{4(7-p)\over (5-p)}\over (r~\ms)^{7-p}}\sim 
{\l^{{2(9-p)\over (5-p)}}\over g_{YM}^2}
\left({\ms\over \mpl}\right)^4
{(T~\ls)^{{4(7-p)\over (5-p)}}\over (r~\ms)^{7-p}}
\ee
The supersymmetry breaking scale $M_{susy}$is identified with the 
temperature T. We will take it to be $T\sim {\cal O}(10^3)$ GeV.
Taking into account (\ref{con}) we can obtain the following bound
\be
\xi_4>> \xi_0={\l^{{2(9-p)\over (5-p)}}\over \gs^{(7-p)\over
(9-p)}~g_{YM}^{4{(8-p)\over (9-p)}}}
\left({M_{susy}\over \ms}\right)^{4(7-p)\over (5-p)}
\left({\ms\over \mpl}\right)^{4+2{(7-p)\over (9-p)}}
\ee
This gives a lower (order of magnitude bound) on the induced cosmological 
constant, namely $\xi_0$.
One can make the cosmological constant two or more orders of magnitude bigger than the bound.
For the cases of interest we have 
\be
\xi_0(p=3)={\l^6\over \gs^{1/2}g_{YM}^{10/3}}\left({M_{susy}\over \mpl}\right)^{
8}
\left({\ms\over \mpl}\right)^{-8/3}
\ee
\be
\xi_0(p=4)={\l^{10}\over \gs^{2/3}g_{YM}^{3}}\left({M_{susy}\over \mpl}\right)^
{12}
\left({\ms\over \mpl}\right)^{-34/5}
\ee
Moreover $\xi_0$ becomes smaller when the string scale is of the same order as the Planck scale. Moreover $g_{YM}\sim {\cal O}(1)$ and we will drop it from the
formula.
Putting in numbers we obtain 
\be
\xi_0(p=3)=10^{-128}~~{\l^6\over \gs^{1/2}}\;\;\;,\;\;\; \xi_0(p=4)=10^{-192}~~
{\l^{10}\over \gs^{2/3}}
\ee
The cosmological constant can be made smaller by making the string couplings
$\gs<<1$, larger by making $N>>1$, or by making $r/\ls$ arbitrarily small.
In particular it should be noted that the cosmological constant can obtain
easily a value close to $10^{-120}$, the favorite number today.
Moreover this can be accomodated even for $\ms\sim M_{susy}$ at the expense of small values of $\gs$.

As argued above it is expected that the extra contributions to 
the vacuum energy
on the universe brane due to loops of brane fields will have extra
suppression factors of the coupling constant.
Moreover, there is an indication \cite{wir} that even ${\cal N}=2$
supersymmetry on the branes might be enough to control in a
similar fashion the vacuum energy in this context.
We can investigate the situation where the stack of N branes are located 
at an orbifold sigularity and have reduced supersymmetry \cite{ks}.
For a smooth such configurations the metric and five-form field have been obtained in \cite{keh}.
The five-form field stregth is not modified whereas the only modification
to the metric is in the geometry of the space that replaces $S^5$.
Their black analogs can be easily written as in (\ref{dmet}) with p=3
where the metric of the five-sphere is replaced by the metric of a U(1) 
bundle on a K\"ahler-Einstein space \cite{keh}. The field stregth of the U(1) connection is equal to the K\"ahler two-form.
Going through the same procedure as before we find the same potential as
in the N=4 case.
This strongly suggests that the supression of the cosmological constant
is due to N=1 rather than N=4 supersymmetry.
The  observations above  may be important for controlling the scale of
the cosmological constant of our universe after supersymmetry breaking.

The second issue to be commented upon concerns the fact that the
effective velocity of light on the probe brane is field-dependent.
This is due here to the black nature of the spectator branes.

A variable velocity of light has a similar effect for the evolution of
the universe
as that due to inflation.
This possibility has been investigated recently \cite{varc} and might
provide
a viable alternative to inflation.
One of the main problems in such an approach is to find a natural
dynamical
evolution of the velocity of light, rather than an ad hoc variability
and to have a certain predictivity on the nature of interactions.

Our example (although rather a toy example) provides a concrete
dynamical framework
for a variable speed of light.
The proposal here is that our brane-universe is the probe brane falling
towards the black brane.
Due to the fact that the effective velocity of light is distance
dependent
as in (\ref{vl}), it becomes smaller with time passing.
Moreover, it induces a Robertson-Walker type of metric on the probe
brane.
This setup deserves further study in order to investigate the
possibility
of a concrete and realistic alternative to inflation.

\vskip 1cm
\centerline{\large \bf Acknowledgments}
\vskip .6cm

I would like to thank C. Bachas and T. Taylor for participating in
early stages of this work and for many discussions.
I would also like to thank A. Kehagias, C. Kounnas, 
M. Laine and J. Wirstam for helpful exchanges.
I am indebted to the Ecole Polytechnique for hospitality where
this work started and the CERN Theory Division for
hospitality during the last phase of this work.

\vskip 3cm
\newpage

\begin{flushleft}
{\Large\bf Appendix A: Calculation of the Dp-brane RR background field
strength}\end{flushleft}
\renewcommand{\theequation}{A.\arabic{equation}}
\renewcommand{\thesection}{A.}
\setcounter{equation}{0}
\def\p{\partial_{r}}

In the string frame the RR field strength satisfies
\be
\nabla^{\mu}F_{\mu\nu_1\cdots\nu_{p+1}}=0\ .
\label{a1}\ee
This is also implied for $p=3$ by the self-duality condition
(\ref{self}).

We parametrize the metric as
\be
ds^2=ds^2_{\ads}+ds^2_{S}\label{amet}
\ee
$$
ds^2_{\ads}=g_{00}(r)dt^2+g_{rr}(r)dr^2+\sum_{i=1}^p
g_{i}(r)dx_i^2
$$
$$
ds^2_S=g_{S}(r)h_{\alpha\beta}dy^{\alpha}dy^{\beta}
$$
where $h_{\alpha\beta}$ is the metric of the unit $S^{8-p}$ sphere.
The only non-zero components are $F_{r012\cdots p}$, except for $p=3$
where the self-duality condition implies also
that $ F_{45678}$ (in the
$S^5$ directions) is non-zero.

Specified to the background of the form (\ref{amet}),
eq.(\ref{a1}) gives
\be
\left[\p-\p\log\left({V_{\ads}\over V_S}\right)\right]F_{r012\cdots
p}=0\ ,
\ee
where $V_{\ads}=\sqrt{g_{00}g_{rr}\prod_{i=1}^p g_i}$ and
$V_S=[g_S(r)]^{(8-p)/2}$.
The solution is
\be
F_{r012\cdots p}=c{V_{AdS}\over V_S}\ ,\label{fr}
\ee
where $c$ does not depend on $r$.
The dual field strength can be obtained using (\ref{self}),\footnote{
We
have used $\epsilon^{012\cdots }=1$}
\be
F^{(p+1)\cdots 8}={c\over V^2_S\sqrt{h}}\;\;\;,\;\;\;F_{(p+1)\cdots
8}=c\sqrt{h}\label{frr}
\ee
The constant $c$ is related to the charge $N$ by the Gauss' law,
\be
\int_{S^{8-p}} F_{(p+1)\cdots 8}=-2\kappa_{10}^2 T_pN\ .
\ee
The coupled $p$-form--gravity field equations impose the relation
eq.(\ref{charge})
between the charge $N$ and the parameters $L$ and $r_0$, so that
\be
c=-{2\kappa_{10}^2 T_p N\over
\Omega_{8-p}}=(p-7)L^{(7-p)/2}\sqrt{r_0^{7-p}+L^{7-p}}
\ee
For the black Dp-brane metric (\ref{dmet}) we obtain
\be
F_{r012\cdots p}=(p-7){L^{(7-p)/2}\sqrt{r_0^{7-p}+L^{7-p}}\over
H_p(r)^2}
\ee
and integrating once and adjusting the constant of integration so that
the p-form falls off at infinity we obtain
\be
C_{012\cdots p}=\sqrt{1+{r_0^{7-p}\over L^{7-p}}}{H_p(r)-1\over
H_p(r)}
\ee
as advertised.


\begin{thebibliography}{99}

\bibitem{mald} J. Maldacena,  Adv. Theor. Math. Phys. {\bf 2} (1998)
231, hep-th/9711200.

\bibitem{gkp} S.S. Gubser, I.R. Klebanov and A.M. Polyakov,
 Phys.
Lett. {\bf B428} (1998) 105,
hep-th/9802109.

\bibitem{w1} E. Witten, Adv. Theor. Math. Phys. {\bf 2} (1998) 253,
hep-th/9802150.


\bibitem{hs} G.T. Horowitz and A. Strominger, \np 360 (1991) 197.

\bibitem{gt} G.W. Gibbons and P.K. Townsend, \prl 71 (1993) 3754,
hep-th/9307049.

\bibitem{pol} J. Polchinski,  Phys. Rev. Lett. {\bf 75} (1995) 4724,
hep-th/9510017.


\bibitem{w2} E. Witten,  Adv. Theor. Math. Phys. {\bf 2} (1998) 505,
hep-th/9803131.

\bibitem{hp} S.W. Hawking and D. Page, Comm.\ Math.\ Phys.\ 87 (1983)
577.



\bibitem{gkpeet} S.S. Gubser, I.R. Klebanov and A.W. Peet, \pr 54
(1996) 3915,  hep-th/9602135.

\bibitem{gkt} S.S. Gubser, I.R. Klebanov and A.A. Tseytlin,
Nucl. Phys. {\bf B534} (1998) 202,
hep-th/9805156.

\bibitem{pt} J. Pawelczyk and S. Theisen, JHEP {\bf 9809} (1998) 010,
hep-th/9808126.

\bibitem{ty} A.A. Tseytlin and S. Yankielowicz,  Nucl. Phys. {\bf B541}
 (1999) 145, hep-th/9809032.

\bibitem{ft} Fotopoulos and T. Taylor, Phys. Rev. {\bf D59} (1999)
061701, hep-th/9811224.




\bibitem{db} J. Polchinski,  hep-th/9611050;\\
C. Bachas, hep-th/9806199; hep-th/9701019.

\bibitem{ds} M. R. Douglas, D. Kabat, P. Pouliot and  S. Shenker
Nucl. Phys. {\bf B485}  (1997) 85,  hep-th/9608024.

\bibitem{li} G. Lifshytz, Phys. Lett. {\bf B388} (1996) 720,
hep-th/9604156.\


\bibitem{doug} M. Douglas ans W. Taylor, hep-th/9807225.

\bibitem{bdhm}
T. Banks, M. Douglas, G. Horowitz and E. Martinec,
hep-th/9808016.


\bibitem{tri}  V. Balasubramanian, P. Kraus, A. Lawrence, S. Trivedi,
  Phys. Rev. {\bf D59} (1999) 104021,
hep-th/9808017;\\
P. Kraus, F. Larsen, S.  Trivedi JHEP {\bf 9903} (1999) 003,
hep-th/9811120.

\bibitem{da} S. Das,  hep-th/9905037.

\bibitem{1} E. Kiritsis and T. Taylor, hep-th/9906048.

\bibitem{mal} J. Maldacena, Phys. Rev. {\bf D57} (1998) 3736,
hep-th/9705053.

\bibitem{gray}
C. G. Callan and J. Maldacena, Nucl. Phys.
{\bf B472} (1996) 591, hep-th/9602043;\\
A. Dhar, G. Mandal and S. R. Wadia, Phys.
Lett. {\bf B388} (1996) 51, hep-th/9605234;\\
S. R. Das, and S. D. Mathur, Nucl. Phys. {\bf B478}
(1996)
 561, hep-th/9606185; Nucl. Phys. {\bf B482} (1996) 153,
 hep-th/9607149;\\
S. Gubser and I. Klebanov, Nucl. Phys. {\bf B482} (1996) 173,
hep-th/9608108;\\
J. Maldacena and A. Strominger, Phys. Rev. {\bf D55}
(1997) 861, hep-th/9609026;\\
S. Gubser and I. Klebanov, Phys. Rev. Lett. {\bf 77} (1996) 4471,
hep-th/9609076

\bibitem{nonr} S. Das,  Nucl. Phys. [Proc.Suppl.] {\bf 68} (1998) 119,
hep-th/9709206.


\bibitem{pol2} A. Peet and J. Polchinski, Phys. Rev. {\bf D59} (1999)
065011, hep-th/9809022.


\bibitem{hw} P. Ho\v rava and E. Witten, Nucl. Phys. {\bf B460}
(1996) 506, hep-th/9510209.

\bibitem{ss} J. Scherk and J. H. Schwarz, Nucl. Phys. {\bf B153} (1979) 61. 


\bibitem{ss2} I. Antoniadis, C. Bachas, D. Lewellen and T. Tomaras,
Phys. Lett. {\bf B207} (1988) 441;\\
C. Kounnas and M. Porrati, Nucl. Phys. {\bf B310} (1988) 355;\\
S. Ferrara, C. Kounnas, M. Porrati and F. Zwirner, Nucl. Phys. {\bf B318} (1989) 75.




\bibitem{varc} J. Moffat, IJMP {\bf D2} (1993) 351, gr-qc/9211020;\\
A. Albrecht and J. Magueijo, Phys Rev. {\bf D59} (1999) 43516,
astro-ph/981101;\\
J. Barrow and J. Magueijo, Phys. Lett. {\bf B447} (1999) 246,
astro-ph/9811073; {\it ibid.}
{\bf B443} (1998) 104, astro-ph/9811072.



\bibitem{wir} J. Wirstam,  hep-th/9902188.

\bibitem{tsen} A. Tseytlin,  Nucl. Phys. {\bf B501} (1997) 41,
hep-th/9701125.

\bibitem{tse} A. Tseytlin,  Nucl. Phys. {\bf B524} (1998) 41,
hep-th/9802133.

\bibitem{the} J. Kapusta, D. Reiss and S. Rudas, Nucl. Phys. {\bf
B263}
 (1986)
207.


\bibitem{bach} C. Bachas,  Phys. Lett. {\bf B374} (1996) 37,
hep-th/9511043;
hep-th/9806199.



\bibitem{fer} S. Ferrara, C. Savoy and L. Girardello, Phys. Lett. {\bf
B105} (1981) 431.

\bibitem{book} E. Kiritsis, ``Introduction to String Theory", Leuven
University
Press, 1998, hep-th/9709062; hep-th/9708130.


\bibitem{brane} V. Rubakov and M. Shaposhnikov,  Phys. Lett. {\bf B125}
(1983) 136.


\bibitem{A} I. Antoniadis, Phys. Lett. {\bf B246} (1990) 377.

\bibitem{L} J.D. Lykken, Phys. Rev. {\bf D54} (1996) 3693,
hep-th/9603133.

\bibitem{ADD} N. Arkani-Hamed, S. Dimopoulos and G. Dvali, Phys.
Lett{\bf B429} (1998) 263, hep-ph/9803315; hep-ph/9807344;\\
I. Antoniadis, N. Arkani-Hamed, S. Dimopoulos and G. Dvali,
Phys. Lett. {\bf B436} (1998) 263, hep-ph/9804398.

\bibitem{DDG} K.R. Dienes, E. Dudas and T. Gherghetta, Phys. Lett. {\bf
B436}(1998) 55, hep-ph/9803466;
Nucl. Phys. {\bf B537} (1999) 47, hep-ph/9806292, hep-ph/9807522;\\
 D. Ghilencea and G.G. Ross, Phys. Lett. {\bf B442} (1998) 165,
hep-ph/9809217;\\
Z. Kakushadze,  Nucl. Phys. {\bf B548} (1999) 205, hep-th/9811193; \\
A. Delgado and M. Quir{\'o}s, hep-ph/9903400;\\
Z. Kakushadze and T.R. Taylor, hep-th/9905137.

\bibitem{ST} G. Shiu and S.-H.H. Tye, Phys. Rev {\bf D58}(1998)
106007, hep-th/9805157\\
Z. Kakushadze and S.-H.H. Tye,  Nucl. Phys. {\bf B548} (1999) 180,
hep-th/9809147.

\bibitem{SR} R. Sundrum,  Phys. Rev. {\bf D59} (1999) 085009,
hep-ph/9805471;
{\it ibid.}  Phys. Rev. {\bf D59} (1999) 085010, hep-ph/9807348;\\
L. Randall and R. Sundrum, hep-th/9810155; hep-ph/9905221;
hep-th/9906064.

\bibitem{B} C. Bachas, JHEP {\bf 9811} (1998) 023, hep-th/9807415.

\bibitem{AB} I. Antoniadis and C. Bachas,  Phys. Lett. {\bf B450}
(1999) 83,
hep-th/9812093.

\bibitem{INT} K. Benakli, hep-ph/9809582;\\
C. Burgess, L.E. Ib{\'a}{\~n}ez and F. Quevedo,  Phys. Lett. {\bf B447}
(1999) 257, hep-ph/9810535;\\
L.E. Ib{\'a}{\~n}ez, C. Mu{\~n}oz and S. Rigolin, hep-ph/9812397.

\bibitem{AP} I. Antoniadis and B. Pioline, Nucl. Phys. {\bf B550}(1999)
41, hep-th/9902055.

\bibitem{Ver} H. Verlinde, hep-th/9906182.

\bibitem{hor} P. Ho\v rava,  Phys. Rev. {\bf D54} (1996) 7561,
hep-th/9608019;\\
I. Antoniadis and M. Quiros, Nucl. Phys. [Proc.Suppl.] {\bf 62} (1998)
312, hep-th/9709023.

\bibitem{ks} S. Kachru ans E. Silverstein, Phys. Rev. Lett. {\bf 80} (1998) 4855; hep-th/9802183.
 
\bibitem{keh} A. Kehagias, Phys. Lett. {\bf B435} (1998) 337, 
hep-th/9805131.

\end{thebibliography}
\end{document}